\documentclass[12pt]{article}

\usepackage{scicite}
\usepackage{graphicx}

\newenvironment{sciabstract}{%
\begin{quote} \bf}
{\end{quote}}

\newcounter{lastnote}
\newenvironment{scilastnote}{%
\setcounter{lastnote}{\value{enumiv}}%
\addtocounter{lastnote}{+1}%
\begin{list}%
{\arabic{lastnote}.}
{\setlength{\leftmargin}{.22in}}
{\setlength{\labelsep}{.5em}}}
{\end{list}}

\title{Dissipationless Quantum Spin Current at Room Temperature}

\author
{Shuichi Murakami,$^{1\ast}$ Naoto Nagaosa,$^{1,2,3}$ Shou-Cheng Zhang$^{4}$\\
\\
\normalsize{$^{1}$Department of Applied Physics, University of Tokyo, Hongo,}
\\
\normalsize{Bunkyo-ku, Tokyo 113-8656, Japan}\\
\normalsize{$^{2}$CERC, AIST Tsukuba Central 4, Tsukuba 305-8562, Japan}\\
\normalsize{$^{3}$CREST, Japan Science and Technology Corporation (JST)}\\
\normalsize{$^{4}$Department of Physics, McCullough Building, Stanford
University,}\\
\normalsize{Stanford CA 94305-4045, USA}\\
\\
\normalsize{$^\ast$To whom correspondence should be addressed; E-mail:  murakami@appi.t.u-tokyo.ac.jp.}
}

\date{}

\begin{document} 




\maketitle

\clearpage 


\begin{sciabstract}
While microscopic laws of physics are invariant under the reversal
of the arrow of time, the transport of energy and information in most
devices is an irreversible process. It is this irreversibility that 
leads to
intrinsic dissipations in electronic devices and limits the
possibility of quantum computation. We theoretically
predict that the electric field can induce a substantial amount of
dissipationless quantum spin current at room temperature, in hole
doped semiconductors such as Si, Ge and GaAs. 
Based on a generalization of the quantum Hall effect,
the predicted effect
leads to efficient spin injection without the need for metallic ferromagnets.
Principles found in this work could enable quantum spintronic
devices with integrated information processing and storage units,
operating with low power consumption and performing reversible
quantum computation.
\end{sciabstract}



Our work is driven by the confluence of the important
technological goals of quantum spintronics  \cite{prinz1998,
wolf2001} with the quest of generalizing the quantum
Hall effect (QHE) to higher dimensions. The QHE is a
manifestation of quantum mechanics observable at
macroscopic scales. In contrast to the 
most transport coefficients in solid state
systems, which are determined by the elastic and the
inelastic scattering rates, 
the Hall conductance $\sigma_H$ in QHE
is quantized and completely independent of any
scattering rates in the system, where 
the transport equation is given by $j_\alpha=\sigma_H
\epsilon_{\alpha\beta} E_\beta$ ( $j_\alpha$ and $E_\beta$
($\alpha,\beta=1,2$) are the charge current and the electric
fields respectively, $\epsilon_{\alpha\beta}$ is the fully
antisymmetric tensor in two dimensions ). 
 While dissipative
transport coefficients are expressed in terms of states in the
vicinity of the Fermi level, the non-dissipative quantum Hall
conductance is expressed in terms of equilibrium response of
all states below the Fermi level. The topological origin of the
QHE is revealed through the fact that the Hall
conductance can also be expressed as the first Chern number of a
$U(1)$ gauge connection defined in momentum space \cite{thouless1982}.
Recently, the QHE has been
generalized to four spatial dimensions\cite{zhang2001A}. In that
case, an electric field $E_\nu$ induces an $SU(2)$ spin current
$j_\mu^i$ ($\mu,\nu=1,2,3,4$, $i=1,2,3$) through the
non-dissipative transport equation $j_\mu^i = \sigma
\eta^i_{\mu\nu} E_\nu$, where $\eta^i_{\mu\nu}$ is the t'Hooft
tensor, explicitly given by
$\eta_{\mu\nu}^i=\epsilon_{i\mu\nu4}+\delta_{i\mu}\delta_{4\nu}-
\delta_{i\nu}\delta_{4\mu}$ and $\sigma$ is a dissipationless
transport coefficient. The quantum Hall response in that system is
physically realized through the spin-orbit coupling in a {\it{time-reversal 
symmetric system}}. At the boundary of this four
dimensional quantum liquid, when both the electric field and the
spin current are restricted to the three dimensional sub-space,
the dissipationless response is given by
\begin{equation}
j_j^i = \sigma_s \epsilon^{ijk} E_k \label{spin_response}
\end{equation}
This fundamental response equation shows that it is possible to
induce a purely topological and dissipationless spin current by an
electric field in the physical, three dimensional space.

We consider a realization of this electric field
induced topological spin current in conventional hole-doped
semiconductors.
In a large class of semiconductors, 
including Si, Ge, GaAs and InSb, the valence bands are four-fold
degenerate at the
$\Gamma$-point (see e.g. Fig.~1).
The
effective Luttinger Hamiltonian~\cite{Luttinger1956A} for holes is given by
\begin{equation}
H_0=\frac{\hbar^{2}}{2m}\left((\gamma_{1}+\frac{5}{2}\gamma_{2})k^{2}-
2\gamma_{2}({\bf k}\cdot{\bf S})^{2}\right), \label{Luttinger}
\end{equation}
where $S_{i}$ is the spin-$3/2$ matrix.
We take the hole picture, and reverse the sign of the energy.
Good quantum numbers for this Hamiltonian are the helicity 
$\lambda=\hbar^{-1}{{\bf k}}\cdot{{\bf S}}/k$,  and the total
angular momentum ${\bf J}=\hbar{\bf x} \times
{\bf k}+{\bf S}$. This kinetic Hamiltonian is diagonalized
in the basis where the helicity operator $\lambda$ is diagonal, and
the eigenvalue is given by
$\epsilon_\lambda({\bf k})=\frac{\hbar^{2}k^{2}}{2m}
(\gamma_{1}+(\frac{5}{2}-2\lambda^{2})\gamma_{2})\equiv\frac{\hbar^{2}
k^{2}}{2m_\lambda}$.
For a given wave vector ${\bf k}$, the Hamiltonian
(\ref{Luttinger}) has two eigenvalues,
$\epsilon_{{\rm H}}({\bf k})=\epsilon_{\lambda=\pm
3/2}({\bf k})=\frac{\gamma_{1}-2\gamma_{2}}{2m}\hbar^{2}k^{2}
\equiv\frac{\hbar^{2}k^{2}}{2m_{\mathrm{H}}}$ and
$\epsilon_{{\rm L}}({\bf k})=\epsilon_{\lambda=\pm
1/2}({\bf k})=\frac{\gamma_{1}+2\gamma_{2}}{2m}
\hbar^{2}k^{2}\equiv\frac{\hbar^{2}
k^{2}}{2m_{{\rm L}}}$, forming Kramers doublets. 
They are referred to as the light-hole (LH)
and heavy-hole (HH) bands.
In semiconductors with zincblende
structure, such as GaAs, inversion symmetry breaking causes an
additional tiny splitting in the LH and HH bands. We can
neglect it when the temperature is much
higher than this splitting. The band structure of
semiconductors deviates from the spherical to the cubic symmetry.
We also neglect this
effect for simplicity, because physics 
described below are not so much affected by it.

We shall consider
the effect of a uniform electric field ${\bf E}$.
Our full
Hamiltonian is thus given by $H=H_0+V({\bf x})$, where
$V({\bf x})=
e{\bf E}\cdot{\bf x}$, and $-e$ is the charge of an
electron.
We assume
that
the
split-off band is
totally occupied.
We first define a $4\times 4$ unitary matrix $U({\bf k})$
which diagonalizes the kinetic Hamiltonian $H_0$. $U({\bf k})$
is defined by $ U ({\bf k})({\bf k}\cdot{\bf S})
U^\dagger({\bf k}) = k S_z$. In the spherical coordinates where
${\bf k}=k(\sin\theta \cos\phi, \sin\theta \sin\phi, \cos\theta)$,
$U({\bf k})$ can be expressed as $U({\bf k})=\exp(i\theta
S_y) \exp(i\phi S_z)$. Under this unitary transformation, the new
Hamiltonian $\tilde H\equiv U({\bf k}) H
U^{\dagger}({\bf k})$ becomes
\begin{equation}
\tilde
H=\frac{\hbar^{2}k^{2}}{2m}\left(\gamma_{1}+\frac{5}{2}\gamma_{2}-
2\gamma_{2} S_z^{2}\right)+ U({\bf k})V({\bf x})U^{\dagger}({\bf k})
\label{Luttinger'}
\end{equation}
Eigenvalues of $S_z$ physically describe the helicity
$\lambda=\hbar^{-1}{\bf k}\cdot{\bf S}/k$ in the original basis. The kinetic
part $H_{0}$ now becomes diagonal, in the
representation where $S_z$ is diagonal. Because ${\bf x}=i\partial_{{\bf k}}$,
the 
potential term becomes $V(\tilde{\bf D})$,
where the covariant derivative $\tilde{\bf D}$ is defined by $\tilde
{\bf D}=i\partial_{\bf k}-\tilde{\bf A}$
and $\tilde{\bf A}=-i U({\bf k})
\partial_{\bf k} U^\dagger({\bf k})$.
As $\tilde{\bf A}$ is a pure gauge potential, there is no
curvature associated with it. Up to this point, the transformation
is exact. We now consider adiabatic transport and make a 
corresponding approximation.
As is usually assumed in the 
transport theory, we neglect the interband transitions,
i.e. the off-block-diagonal matrix elements of $\tilde{{\bf A}}$ 
connecting the LH and HH bands.
Then we arrive at a non-trivial
adiabatic gauge connection ${\bf A}$ (online supporting text), which takes a 
block-diagonal form in the LH and HH subspace. As each
band is two-fold degenerate, the gauge connection is in general
non-Abelian. However, ${\bf A}$ has no matrix elements
connecting the $\lambda=3/2$ and $\lambda=-3/2$ states in the HH band, because
the gauge field $\tilde{{\bf A}}$ only connects states with
helicity difference $\Delta\lambda=0,\pm 1$. Therefore, the
non-Abelian structure is only present in the LH band. For 
simplicity of presentation, we shall first make an additional,
{\it Abelian approximation} (AA), in which only the diagonal components
in ${\bf A}$ are retained. Afterwards,  
we shall give our final results including fully the non-Abelian
corrections.

Within the AA, ${\bf A}$ is a diagonal
$4\times 4$ matrix in the helicity basis.
As a band-touching point acts as a Dirac 
magnetic monopole in momentum space\cite{berry1984},
each diagonal component of ${\bf A}({\bf k})$ is given by that of 
a Dirac monopole at ${\bf k}=0$, with the monopole strength $eg$
given by $\lambda$.
The associated field strength is
given by
\begin{eqnarray}
&&F_{ij} \equiv i[D_i,D_j]
=\epsilon_{ijk}\lambda\frac{
k_{k}}{k^{3}} \label{F}
\end{eqnarray}
The effective
Hamiltonian takes the form
\begin{equation}
H^{{\rm eff}}=\frac{\hbar^{2}k^{2}}{2m_\lambda} + V({\bf x}).
\label{Heff}
\end{equation}
Henceforth, $x_{i}$ denotes  a covariant
derivative in momentum space:
$x_i=D_i=i\partial/\partial{k_i}-A_{i}({\bf k})$.
Note that the definition of $x_{i}$
has changed by projecting the original Hamiltonian $H$ onto 
the HH or LH band.
While $H^{{\rm eff}}$ seems to be trivial, its non-trivial
dynamics is revealed through the non-trivial commutation relations
\begin{equation}
[k_i,k_j]=0 \ ,\ [x_i,k_j]=i \delta_{ij} \ ,\ [x_i,x_j]=-iF_{ij}.
\label{non-commute}
\end{equation}
Such situation also happens in 
the Gutzwiller projection of the SO(5) model \cite{zhang1999}.
It also resembles the non-trivial commutation
relation between the position operators of a
two-dimensional-electron-gas projected onto the
lowest-Landau-level\cite{prangegirvin}, 
where $F_{ij}=B\epsilon_{ij}$, and $B$ is the
external magnetic field. This general algebraic structure, called
``non-commutative geometry", also underlies the four-dimensional
QHE model \cite{zhang2001A}.
In our present context, the non-commutativity between the
three-dimensional coordinates arises from the magnetic monopole in
momentum space, and it is a natural generalization of the QHE
to three dimensions.

The equation of motion for holes can be derived easily from 
Eqs.~\ref{Heff} and \ref{non-commute} as
\begin{equation}
\hbar\dot{k}_{i}=eE_{i}, \ \ \dot{x}_{i}=\frac{\hbar
k_{i}}{m_{\lambda}}+ F_{ij}\dot{k}_{j}. \label{eom}\end{equation} 
The last term, proportional to $F_{ij}$, is a topological term,
describing the effect of the magnetic monopole on the orbital
motion. It represents a ``Lorentz force"
in momentum space, making the hole
velocity non-collinear with its momentum, in contrast to 
the usual situations. 
In fact, if we interchange the roles of $x$ and
$k$ in this term, it becomes the Lorentz force
for a charged particle moving in the presence of a magnetic
monopole in real space. This set of equations can be integrated
analytically (supporting online text), and 
the resulting trajectory is shown in Fig.~2.
The hole motion in real space obtains a shift
perpendicular to ${\bf S}$. This shift is analogous to the
deflection of a charged particle by a magnetic monopole, in a
direction perpendicular to the plane spanned by its position and
velocity vectors \cite{Jackson}. It 
causes a spin current perpendicular to both ${\bf E}$ and
${\bf S}$. 
For example, for ${\bf E}$  parallel to
the $+z$ direction, 
the spin current for each band at zero
temperature, with spin parallel to the $x$ axis, flowing to the
$y$ direction is given by
\begin{equation}
j_{y}^{x}
=\frac{eE_{z}}{36\pi^{2}}(9k_{F}^{{\rm H}}+k_{F}^{{\rm L}}),
\label{abelian-spin-current}
\end{equation}
which is obtained by summing contributions from all the filled states.
Here we assumed
that the equilibrium momentum distribution is attained by the
random impurity scattering which causes the charge relaxation.
Note that Eq.~\ref{eom} describes only the ballistic motion and scattering
by random impurities would lead to additional contributions to the
spin current. As one can see from the detailed discussions in the
supporting online text, these extrinsic effects are not only
small, but also scale with a higher power of $k_F \sim n^{1/3}$,
where $n$ is the hole density. Therefore, by plotting
$\sigma_s/n^{1/3}$ against $n$, and extrapolating to the limit of
$n \rightarrow 0$, the constant intercept would uniquely determine
our predicted dissipationless spin conductivity.


It is worth noting that this AA becomes exact in  
zero-gap semiconductors, e.g. $\alpha$-Sn. In this class of materials,
the bottom of the conduction band and the top of the valence band 
correspond to the LH and HH bands in other semiconductors like GaAs.
These two bands touch at ${\bf k}=0$. In this case, p-doping introduces 
holes only into the HH band, and the AA becomes exact.

The electric-field-induced spin current can also be understood in
terms of the conservation of the {\it total} angular momentum
${\bf J}=\hbar{\bf x} \times {\bf k}+{\bf S}$. As
remarked earlier, 
${\bf J}$ commutes with
$H_0$. When ${\bf E}$ is parallel to the $z$-direction, 
${J_z}$ also commutes with the
potential. Therefore, substituting ${\bf S}=\lambda
\hbar{\bf \hat k}= \lambda\hbar{\bf k}/k$, we obtain
\begin{equation}
\dot{J}_z=\hbar(\dot{\bf x} \times {\bf k})_z +
\hbar({\bf x} \times \dot{\bf k})_z
+\lambda\hbar \dot{\bf \hat k}_z=0 
\label{conservation}
\end{equation}
The second term, representing the torque,
vanishes in our case since $\dot{\bf k}$ points along the
$z$ direction. The first term 
$\hbar(\dot{\bf x} \times {\bf k})_z$ vanishes in usual
problems; however, it does not in our case, due to the
non-collinearity of the velocity and the momentum. Furthermore,
the first
term, describing the time derivative of the orbital angular
momentum ${\bf L}=\hbar{\bf x}\times{\bf k}$, 
is proportional to the spin current. The third
term  $\lambda \hbar\dot{\bf \hat k}_z$
describing the time derivative of the spin angular momentum ${\bf S}$,
can be easily evaluated from the acceleration equation in Eq.~\ref{eom}. 
Therefore,
we see that the conservation of the total angular momentum
Eq. \ref{conservation} directly implies the spin current 
Eq.~\ref{abelian-spin-current}.
The spin current flows in such a way
that the change of ${\bf L}$ exactly cancels the
change of ${\bf S}$.

We now discuss the correction due to the
non-Abelian nature of the gauge connection of the LH band. 
Remarkably, even
though the gauge connection is non-Abelian, the associated field
strength is Abelian, and gives a correction factor of $(-3)$,
compared to the AA \cite{zee1988,arovas1998A,jungwirth2002}.
The equation of motion modified accordingly agrees with that obtained by
generalizing the wave packet formalism \cite{sundaram1999}
to the non-Abelian case. 
This non-Abelian correction gives the following
result for the spin current:
\begin{equation}
j_{y}^{x} =
\frac{eE_{z}}{12\pi^{2}}(3k_{F}^{{\rm H}}-k_{F}^{{\rm L}}) =
\frac{\hbar}{2e} \sigma_s E_z. \label{spin-current}
\end{equation}
Here we  defined $\sigma_s$ to have the same dimension as the
electrical conductivity, to facilitate comparison. The spin
current equation is rotationally invariant, with the
covariant form given in Eq.~\ref{spin_response}, and is the central
result of our paper.
In contrast with similar effects 
\cite{hirsch1999,bulgakov1999},
this spin current has a topological character; the spin
conductivity $\sigma_s$ in Eq.~\ref{spin_response} is independent of
the mean free path and relaxational 
rates,
and all states below the Fermi energy contribute to the
spin current, where each contribution is purely determined by the 
gauge curvature in momentum space, similar to the
QHE \cite{thouless1982}.
Assuming
the hole density $n = 10^{19}$cm$^{-3}$, the mobility of the
holes at room temperature in GaAs is $\mu = 50$cm$^2$/Vsec \cite{lb},
and the conductivity is $\sigma = e n \mu = 80 \Omega^{-1}$cm$^{-1}$.
On the other hand, the spin Hall conductivity $\sigma_s$ in
Eq.~\ref{spin-current} is estimated as $\sigma_s 
\sim 80 \Omega^{-1} {{\rm cm}}^{-1}$, being of the same order with $\sigma$. 
For lower carrier concentration,
$\sigma_{s}$ becomes larger than $\sigma$; for
$n=10^{16}{\rm cm}^{-3}$, we have
$\sigma=0.6\Omega^{-1}{\rm cm}^{-1}$ and
$\sigma_s=7\Omega^{-1}{\rm cm}^{-1}$.
At finite temperature, Eq.~\ref{spin-current} 
is modified only through the Fermi distribution
function $n^{\lambda}({\bf k})$.
Since the typical energy difference between the LH and HH bands at the same 
wavenumber is about 0.1eV, which largely exceeds the energy scale of the room
temperature $\sim$ 0.025eV, our predicted effect remains of the same order
even at room temperature.

We remark that the non-dissipative spin transport equation
Eq. \ref{spin_response} does not violate the time-reversal symmetry
${\cal T}$. Our microscopic Hamiltonian $H$, the electric field
${\bf E}$ and the spin current are all ${\cal T}$ invariant. Therefore,
the electric field and the spin current can be related by a ${\cal
T}$ symmetric, dissipationless transport coefficient $\sigma_s$.
This situation is to be contrasted with the Ohm's law. As the
charge current is odd under ${\cal T}$, while the electric field
is even, they can only be related by a ${\cal T}$ antisymmetric,
dissipative transport coefficient, namely the charge conductivity.
One of the main objective in quantum computing is to achieve {\it
reversible computation} \cite{feynman,loss1998}. From the above
analysis, we see that there is a fundamental difference between
the ordinary irreversible electronics computation based on the
Ohm's law, and the reversible spintronics computation based on
Eq.~\ref{spin_response}. The time reversal symmetry
property encoded in Eq.~\ref{spin_response}
could provide a fundamental principle for the reversible quantum
computation.

This spin current is also useful for spin injection into
semiconductors. While effective spin injection is necessary for
spintronic devices, it has been an elusive issue
\cite{wolf2001}. Usage of ferromagnetic metals is not
practical because most of the spin polarizations are lost at the
interface due to conductivity mismatch between metal and
semiconductor\cite{hammar1999,schmidt2000}. Spin injection from
ferromagnetic semiconductors such as Ga$_{1-x}$Mn$_{x}$As has been
successful \cite{ohno1999b,fiederling1999,mattana2003}.
Nevertheless, $T_c$ is at most 110K for Ga$_{1-x}$Mn$_{x}$As,
still too low for practical use at room temperature. Thus it is
desirable to find an effective method for spin injection. The
electric-field-induced spin current serves as a
spin injector, because it creates a spin current {\it inside the
semiconductor}. One might worry that the short relaxation time $\tau_s
\sim 100 {\rm fsec}$ \cite{hilton2002} of the hole spins.
This shortness of $\tau_{s}$ is because
the strong spin-orbit interaction in the valence band
combines the relaxation of momenta and spins \cite{opticalorientation}.
Actually most of the efforts on the
spintronics in GaAs have been focused on electrons in the
conduction bands, having much longer spin-relaxation time 
($\sim 100 {\rm psec}$) \cite{ohno1999a}.
Nevertheless, our spin
current is free from such rapid relaxation of spins of holes, because
it is a purely quantum mechamical effect with
equilibrium spin/momentum distribution.
Only when the spin/momentum distribution
deviates from equilibrium, e.g., in 
spin accumulation at boundaries of the sample,
the rapid relaxation of hole spins becomes effective.

One can consider following experimetal setups for 
detection of the constant spin supply from  p-GaAs.  When the
electric field is applied along the $z$-direction and the electric
current $J_z$ is induced, the $s^{x}$-spin current $j_{y}^{x}$ will
flow along the $y$ direction. One possibility is to see the
spin-dependent electric transport through a ferromagnetic
electrode with the magnetization ${\bf M}$ along $\pm
x$-direction attached to the positive-$y$ side of the sample. With
a lead connecting this electrode and the other  (negative-$y$)
side of p-GaAs as shown in Fig.~3{\bf A}, 
one should see a change of the electric current $I$
depending on the direction of
${\bf M}$. 
The ratio of $I$ when ${\bf M}$ is along $\pm x$-direction, $I(+x)/I(-x)$, 
is expected to 
be well larger than unity.
For the ferromagnetic electrode, ferromagnetic
metals are not efficient, because of the conductance mismatch
\cite{schmidt2000}. 
Instead, ferromagnetic
semiconductors 
will be suitable.
Another possibility is to measure circular 
polarization of light emitted via recombination
with electrons. This can be achieved by a similar experimental
setup in Ref.~\cite{ohno1999b}, replacing
(Ga,Mn)As by p-GaAs, where the quantum well structure of (In,Ga)As
is sandwiched by p-GaAs and n-GaAs (Fig.~3{\bf B}). 
The spin
current injected along the $y$-direction will be recombined with
the electrons supplied from the n-GaAs in the (In,Ga)As quantum
well. 

When the system is not connected to the leads along the $y$-direction,
spins  accumulate near the edges  of the
sample. This spin polarization can in principle be measured
by the Kerr rotation.
The spin distribution is determined by a balance between 
the spin current supply and the spin relaxation.
At room temperature, because $\tau_s= 100{\rm fsec}$ \cite{hilton2002}
is rather short, 
the area density of spin accumulation at the sample surface,
$j_{y}^{x}\tau_{s}$, is too small to be observed.
However, there are several ways to make $\tau_{s}$ longer.
One is to lower the temperature. 
Another is to inject the spin into
n-GaAs through the p-n junction as
demonstrated recently \cite{johnstonhalperin2002,
kohda2003}.  
The spin lifetime of electrons in
GaAs is $100$psec,  
10$^3$ times longer than that of holes.
Thereby the spin current can be detected by the Kerr rotation
through surface reflection.



\begin{thebibliography}{99}
\bibitem{zee1988}
   A. Zee, {\it Phys. Rev. A} {\bf 38}, 1 (1988). 
Note that there is a misprint in this reference;
in the left column of page 2, the field strength
should be $F=imd\Omega$.
\end{thebibliography}



\begin{scilastnote}
\item
We thank C. Herring and M. Gonokami for helpful discussions. This work is
supported by
Grant-in-Aids
from the Ministry of Education, Culture, Sports, Science and Technology
of Japan,
the NSF under grant numbers DMR-9814289, and the US
Department of Energy, Office of Basic Energy Sciences under
contract DE-AC03-76SF00515.
\end{scilastnote}




\clearpage
\begin{figure}[h]
\begin{center}
\includegraphics[scale=0.7]{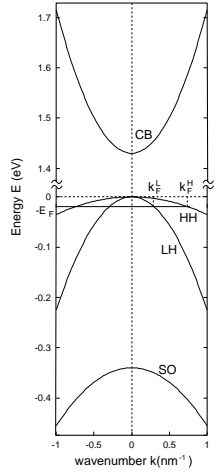}
\caption{
Approximate band structure
of GaAs. We neglect the small
splitting due to inversion symmetry breaking. 
We also neglect the anisotropy of the bands.
The conduction band (CB) is two-fold degenerate. The valence band 
consists of the
heavy hole (HH),
the light hole (LH), and the split-off (SO) bands, each of which is two-fold 
degenerate. We consider the p-GaAs, and the Fermi momentum for
each band is labelled as $k_F^{{\rm H}}$ and $k_F^{{\rm L}}$ respectively.
The Fermi energy shown in the figure corresponds to $n=10^{19}{\rm cm}^{-3}$.
}
\end{center}
\end{figure}

\begin{figure}[h]
\begin{center}
\includegraphics[scale=0.7]{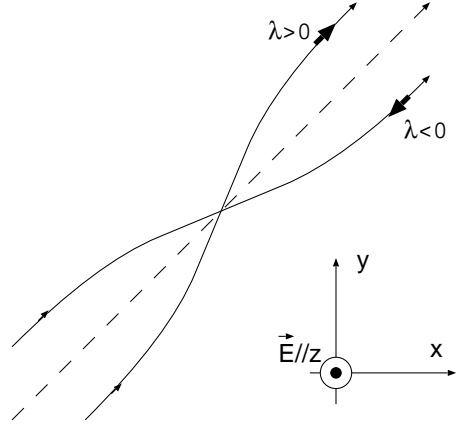}
\end{center}
\caption{
The
real-space
trajectory of the hole obtained by solving
Eq.(\ref{eom}),
The electric field ${\bf E}=e^{-1}{\bf \nabla}V$ is parallel to the $+z$ 
direction.
Due to the non-commutative relation between the
components of the position operator $x_i$, i.e., $[x_i,x_j] = 
-iF_{ij}$ with $F_{ij}$ being the gauge curvature defined in the
momentum space, the hole obtains a transverse velocity whose
direction depends on the helicity $\lambda = {\bf k} \cdot {\bf S}
/k$ indicated by the thick arrow. The broken line is parallel to
${\bf k}$.}
\end{figure}

\begin{figure}[h]
\begin{center}
\includegraphics[scale=0.7]{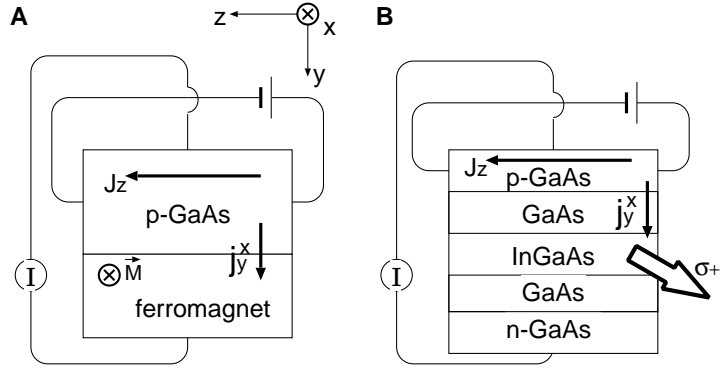}
\caption{
Experimental
setups for the detection of spin current induced by an electric field.
{\bf (A)} Detection by attaching a ferromagnetic electrode.  The
dependence of current $I$ flowing into the electrode on the direction
of the magnetization ${\bf M}$ is
to be measured. {\bf (B)} Detection by measuring the polarization of
the emitted light in the quantum well of (In,Ga)As.
Right and left circular polarization will be switched when the direction
of the external electric field is reversed.}
\end{center}
\end{figure}

\clearpage

\setcounter{equation}{0}
\renewcommand{\theequation}{S\arabic{equation}}
\renewcommand{\theenumi}{S\arabic{enumi}}

\begin{center}
{\Large 
Supporting online material
for the article ``Dissipationless Quantum Spin Current at Room Temperature''}
\end{center}

\vspace{5mm}

\subsection*{Luttinger Hamiltonian for semiconductors}

Here we briefly summarize basic aspects of the Luttinger Hamiltonian
for semiconductors.
In isolated
atoms, spin-orbit coupling leads to a splitting of the p orbitals
into four-fold degenerate $P_{3/2}$ and two-fold
degenerate $P_{1/2}$ levels. In a large class of semiconductors, 
including Si, Ge, GaAs and InSb, the $P_{3/2}$ levels form
the top of the valence bands, which are separated from $S$-like
conduction bands. Therefore, the valence bands are four-fold
degenerate and the conduction bands are two-fold degenerate at the
$\Gamma$-point (see e.g. Fig.~1).
The form of the effective Hamiltonian including
spin-orbit coupling can be determined from symmetry arguments
alone.  In crystals with inversion symmetry, band energies depend
quadratically on the momentum away from the center; therefore, a
rotationally invariant Hamiltonian can only contain two possible
terms, ${\bf k}^2$ and $({\bf k}\cdot {\bf S})^2$, where ${\bf S}$
is the $2\times 2$ Pauli matrix for the two-fold conduction band
and the $4\times 4$ spin matrix for the four-fold valence band.
As a square of each Pauli matrix is the identity matrix, to
the lowest order, there is no spin-orbit coupling in the
conduction band with inversion symmetry. In the presence of
inversion symmetry breaking, either due to intrinsic crystal
symmetry or in hetero-junction systems, the Dresselhaus and the
Rashba terms for spin-orbit coupling are possible
({\it S1--S3\/}). On the other hand,
in the four-fold valence band, spin-orbit coupling affects the
band structure, even without inversion symmetry. As the square
of each component $S_{i}$  $(i=1,2,3)$ for $S=3/2$ is non-trivial, 
these two terms combine to form the
effective Luttinger Hamiltonian ({\it S4\/}) for holes:
\begin{equation}
H_0=\frac{\hbar^{2}}{2m}\left((\gamma_{1}+\frac{5}{2}\gamma_{2})k^{2}-
2\gamma_{2}({\bf k}\cdot{\bf S})^{2}\right), \label{Luttinger-SOM}
\end{equation}
where we take the hole picture, and reverse the sign of the energy.
The eigenvalues are classified by the helicity $\lambda=
\hbar^{-1}
{\bf k}\cdot{\bf S}/k$; $\lambda=\pm 1/2$ corresponds to the LH band, 
and $\lambda=\pm 3/2$ to the HH band.

\subsection*{Details for calculation of ${\bf A}({\bf k})$}

The spin-$3/2$ matrices $S_{i}$ are written as
\begin{equation}
S_{x}=\left(
\matrix{
&\frac{\sqrt{3}}{2}&&\cr
\frac{\sqrt{3}}{2}&&1&\cr
&1&&\frac{\sqrt{3}}{2}\cr
&&\frac{\sqrt{3}}{2}&\cr
}\right),\
S_{y}=\left(
\matrix{
&-\frac{\sqrt{3}}{2}i&&\cr
\frac{\sqrt{3}}{2}i&&-i&\cr
&i&&-\frac{\sqrt{3}}{2}i\cr
&&\frac{\sqrt{3}}{2}i&\cr
}
\right),\
S_{z}=\left(
\matrix{
\frac{3}{2}&&&\cr
&\frac{1}{2}&&\cr
&&-\frac{1}{2}&\cr
&&&-\frac{3}{2}\cr
}
\right).
\end{equation}
By using $U({\bf k})=\exp(i\theta
S_y) \exp(i\phi S_z)$,
the gauge field $\tilde{\bf A}=-i U({\bf k})
\partial_{\bf k} U^\dagger({\bf k})$ is calculated as 
\begin{equation}
\tilde{{\bf A}}\cdot d{\bf k}=\left(
\matrix{
-\frac{3}{2}\cos\theta d\varphi &
\frac{\sqrt{3}}{2}(\sin \theta d\varphi+id\theta)
&&\cr
\frac{\sqrt{3}}{2}(\sin \theta d\varphi-id\theta)
&-\frac{1}{2}\cos\theta d\varphi &
\sin\theta d\varphi+id\theta
\cr
&\sin\theta d\varphi -id\theta&
\frac{1}{2}\cos\theta d\varphi &
\frac{\sqrt{3}}{2}(\sin \theta d\varphi+id\theta)
\cr
&&\frac{\sqrt{3}}{2}(\sin \theta d\varphi-id\theta)
&\frac{3}{2}\cos\theta d\varphi \cr
}
\right),
\end{equation}
where the first and fourth columns correspond to the HH, while
the second and third ones correspond to the LH.
As $\tilde{\bf A}$ is a pure gauge potential, its curvature $F_{ij}
=i[D_{i},D_{j}]$ vanishes.
In considering adiabatic transport, we neglect the interband transitions,
i.e. the off-block-diagonal matrix elements of $\tilde{{\bf A}}$ 
between the LH and HH bands.
Then we arrive at a non-trivial
adiabatic gauge connection 
\begin{equation}
{\bf A}'\cdot d{\bf k}=\left(
\matrix{
-\frac{3}{2}\cos\theta d\varphi &
&&\cr
&-\frac{1}{2}\cos\theta d\varphi &
\sin\theta d\varphi+id\theta
&\cr
&\sin\theta d\varphi -id\theta&
\frac{1}{2}\cos\theta d\varphi &
\cr
&&
&\frac{3}{2}\cos\theta d\varphi \cr}
\right),
\end{equation}
which takes a 
block-diagonal form in the LH and HH subspace. As the states in each
band are two-fold degenerate, the gauge connection is in general
non-Abelian. However, 
the
non-Abelian structure is only present in the LH band;
namely, in the LH subspace the matrix ${\bf A}'$ has off-diagonal 
components, while in the HH subspace ${\bf A}'$ is diagonal.
This is because 
the gauge field $\tilde{{\bf A}}$ only connects states with
helicity difference $\Delta\lambda=0,\pm 1$, 
and does not connect $\lambda=\pm 3/2$
states in the HH band.
By employing the Abelian approximation (AA) 
we neglect the off-diagonal elements of ${\bf A}'$,
and we get thereby
\begin{equation}
{{\bf A}'}_{\rm Abelian}=-S_{z}\cos\theta \nabla\phi,
\end{equation}
which is diagonal in the basis where $S_{z}$ is diagonal ({\it S5\/}). 
As has been recognized in ({\it S6\/}),
the degeneracy point acts as the source of the gauge field, i.e. Dirac 
magnetic monopole.
The present form of ${\bf A}'$ corresponds to a Dirac monopole at
the origin of the momentum space, with the monopole strength $eg$
given by $S_{z}$, in accordance with ({\it S6\/}). 
To return to the original basis where the helicity $\lambda$ becomes diagonal,
we make the gauge transformation
\begin{equation}
{\bf A}_{{\rm Abelian}}({\bf k})=U^{\dagger}({\bf k}){{\bf A}'}_{\rm Abelian}
({\bf k})
U({\bf k})+i\frac{\partial U^{\dagger}({\bf k})}{\partial {\bf k}}
U({\bf k})
\end{equation}
The monopole strength is then expressed as $eg=\lambda$.
The associated magnetic field strength is
given by
\begin{equation}
F_{ij} \equiv i[D_i,D_j]
=\epsilon_{ijk}\lambda\frac{
k_{k}}{k^{3}} \label{F-SOM}
\end{equation}

We now discuss the correction due to the
non-Abelian nature of the gauge connection of the LH band. 
Remarkably, even
though the gauge connection is non-Abelian, the associated field
strength is Abelian, and is 
given by ({\it S5\/})
\begin{equation}
F_{ij} =\epsilon_{ijk}\lambda\left( 2\lambda^2-\frac{7}{2}\right) \frac{
k_{k}}{k^{3}}. \label{F-correction}
\end{equation}
This has a correction factor of $(-3)$ only in the LH
compared to the AA.
As the spin current depends only on the field strength in
momentum space, the spin current acquires an extra factor of $(-3)$ only 
in the LH, compared with that within the AA.

\subsection*{Real-space trajectory of the hole motion}

The equation of motion (Eq.~7) within the AA can be integrated analytically.
When ${\bf E}$ is parallel to
the $+z$ direction, we get
\begin{eqnarray*}
&& k_{x}(t) =k_{x0},\ k_{y}(t) =k_{y0},\ k_{z}(t)=k_{z0}+eE_{z}t/\hbar,\\
&& z(t)=z_{0}+\frac{\hbar k_{z0}}{m_{\lambda}}t+\frac{eE_{z}}
{2m_{\lambda}}t^{2},
\\
&&x(t)=x_{0}+\frac{\hbar k_{x0}}{m_{\lambda}}t
-\frac{\lambda k_{y0}}{k_{x0}^{2}+k_{y0}^{2}}
\frac{eE_{z}t/\hbar
+k_{z0}}{\sqrt{k_{x0}^{2}+k_{y0}^{2}+(eE_{z}t/\hbar+k_{z0})^{2}}},\\
&&y(t)=y_{0}+\frac{\hbar k_{y0}}{m_{\lambda}}t
+\frac{\lambda k_{x0}}{k_{x0}^{2}+k_{y0}^{2}}
\frac{eE_{z}t/\hbar+k_{z0}}{\sqrt{k_{x0}^{2}
+k_{y0}^{2}+(eE_{z}t/\hbar+k_{z0})^{2}}}.
\end{eqnarray*}
The last terms in $x(t)$ and $y(t)$ represent a shift of the particle
position, in a direction perpendicular to ${\bf k}$, as
presented in Fig.~2. As the directions of the
spin is parallel (antiparallel) to ${\bf k}$ for $\lambda>0$
($\lambda<0$), the hole motion in real space obtains a shift
perpendicular to ${\bf S}$. This shift
causes a spin current perpendicular to both ${\bf E}$ and
${\bf S}$. 

\subsection*{Calculation of spin current}
The motion along the $z$-direction in the equation of motion
(Eq.~7) is free acceleration by the electric field. In reality,
because this acceleration is suppressed by random scattering we
should take into account the charge relaxation by random
scattering as is discussed in details in the next section.
For example, the spin current for each band at zero
temperature, with spin parallel to the $x$ axis, flowing to the
$y$ direction is given by
\begin{eqnarray}
&&j_{y}^{x{\rm H}}=\frac{\hbar}{3}\sum_{\lambda=\pm\frac{3}{2},
{\bf k}}\dot{y}
\frac{\lambda k_{x}}{k} n^{\lambda}({\bf k})
=\frac{3e}{2}E_{z}\sum_{{\bf k}}\frac{ 
n^{{\rm H}}({\bf k})
k_{x}^{2}}{k^{4}}
=\frac{eE_{z}k_{F}^{{\rm H}}}{4\pi^{2}},
\label{abelian-spin-current-H}
\\
&&j_{y}^{x{\rm L}}=\frac{\hbar}{3}\sum_{\lambda=\pm\frac{1}{2},
{\bf k}}\dot{y} \frac{\lambda k_{x}}{k} n^{\lambda}({\bf k})
=\frac{e}{6}E_{z}\sum_{{\bf k}}\frac{ 
n^{{\rm L}}({\bf k})
k_{x}^{2}}{k^{4}} = \frac{eE_{z}k_{F}^{{\rm L}}}{36\pi^{2}},
\label{abelian-spin-current-L}
\end{eqnarray}
where $n^{\lambda}({\bf k})$ is a filling of holes in the band
with helicity $\lambda$; 
$n^{\pm 3/2}({\bf k})=n^{{\rm H}}({\bf k})$ and 
$n^{\pm 1/2}({\bf k})=n^{{\rm L}}({\bf k})$ 
are fillings for the HH and LH bands,
respectively. Charge relaxation is already included in 
this result by putting the Fermi surfaces as in their equilibrium 
position $|{\bf k}|=k_{F}$.
We have taken into account that the expectation
value of the electron spin ${\bf s}$ is one-third of that of
${\bf S}$: ${\bf s}=\frac{1}{3}{\bf S}$. This is because
the spin-3/2 matrix ${\bf S}$ is a summation of the spin
angular momentum ${\bf s}$ with spin one-half and the atomic
orbital angular momentum ${\bf l}$ with spin one
({\it S4\/}).

By including the non-Abelian correction in the LH band 
(Eq.~\ref{F-correction}), 
the spin current is given by
\begin{eqnarray}
j_{y}^{x} &=&\frac{\hbar}{3}
{\rm tr}
\sum_{{\bf k}}\dot{y}^{n}({\bf k})
S_{x}
n^{\lambda}({\bf k})
=\frac{eE_{z}}{3}{\rm tr}\sum_{{\bf k}}F_{yz}({\bf k})
S_{x}n^{\lambda}
({\bf k})\nonumber \\
&=&\frac{eE_{z}}{3}
{\rm tr}
\sum_{{\bf k}}\frac{k_{x}S_{x}}{k^{3}}\lambda
\left(2\lambda^{2}-\frac{7}{2}\right)
n^{\lambda}({\bf k})
=\frac{eE_{z}}{9}
\sum_{{\bf k},\lambda}\frac{1}{k^{2}}\lambda^{2}
\left(2\lambda^{2}-\frac{7}{2}\right)
n^{\lambda}({\bf k})
\nonumber \\ &=&
\frac{eE_{z}}{12\pi^{2}}(3k_{F}^{{\rm H}}-k_{F}^{{\rm L}}).
\label{spin-current-SOM}
\end{eqnarray}
Interestingly, in the Ioffe-Regel limit of $k_F l \sim 1$
the size of the spin conductivity is comparable to that of
the charge conductivity $\sigma\sim k_F^2 l$.

The formula, Eq.~\ref{spin-current-SOM}, is at zero temperature;
for finite temperature, it is modified only through the Fermi distribution
function $n^{\lambda}({\bf k})$.
For example, at room temperature and $n=10^{19}{\rm cm}^{-3}$,
the nominal value of the energy difference between the LH and HH bands
at the same wavenumber is 0.1eV, and it
largely exceeds both the
temperature ($\sim 0.025$eV) and the energy scale for momentum relaxation
$\hbar/\tau_{p}\sim0.006$eV, 
where we estimated the momentum relaxation time $\tau_{p}$ to be of the 
same order ({\it S7\/}) as 
the spin relaxation time for holes $\tau_{s}\sim 100{\rm fsec}$
({\it S8\/}).
Thus the value of the spin current remains of the same order as in the
zero temperature.

\subsection*{Impurity scattering and relaxation}
Our equation of motion Eq.~7 describes only the ballistic motion,
and the relaxation process due to the random scattering by
impurities and phonons are not explicitly taken into account. The
relaxation is essential to attain the thermal equilibrium, which
we assumed to obtain Eq.~10. Our result is similar to the
Karplus-Luttinger (KL) term ({\it S9\/}) in the context of the
anomalous Hall effect (AHE). The effects of the random scatterings
have been discussed in the context of AHE by many authors
({\it S10--S14\/}); here we shall use some
of these results to estimate the impurity contribution to the spin
current. Kohn-Luttinger ({\it S11\/}) and Luttinger
({\it S12\/}) derived the formula for AHE based on the
rigorous treatment of the density matrix in the expansion of the
impurity scattering strength $v$, reproducing all the
contributions to $\sigma_{xy}$ up the $v^0$ order. The skew
scattering contribution proposed by Smit ({\it S10\/}) is of the
order of $v^{-1}$, and hence is expected to be important in the
weak disorder case. However it is proportional to $\langle V^3 
\rangle_c$ ($V$
is the impurity potential, and $\langle\rangle_c$ 
is the cumulant average) and hence vanishes for the Gaussian
distribution of the disorder potential. Furthermore, this skew
scattering mechanism predicts $\rho_H \propto \rho$ ($\rho$: Hall
resistibity, $\rho$: diagonal resistivity), which is not usually
observed experimentally including the doped GaAs. The contribution
of the order of $v^0$ contains two terms, both of which are
independent of the scattering. One is the KL term, which is
similar to our dissipationless spin current, and the other term is
later interpreted by Berger ({\it S13\/}) as the side-jump
contribution accompanied by the impurity scattering. In the
following, we shall estimate their relative magnitudes and point
out the important fact that they scale differently with the hole
density $n$, and can therefore be uniquely distinguished in the
low density limit.

In the present case of GaAs, there are several evidences that the
intrinsic AHE (KL term) is the dominant contribution. Jungwirth
et al.~({\it S15\/}) showed that KL term explains the experimentally
observed $\sigma_{xy}$ quantitatively for ferromagnetic (Ga,Mn)As
and (In,Mn)As. Especially the difference between (Ga,Mn)As and
(In,Mn)As is successfully attributed to the different ratio
$m_{LH}/m_{HH}$ of the effective masses of light- and heavy hole
bands. Also one can estimate the contribution of side jump and
skew scattering as follows. By modifying the expression for the AHE
due to side jump mechanism ({\it S13,S14\/}), the side-jump 
contribution to the 
spin Hall
conductance is given by
\begin{equation}
\sigma_{s}^{\rm side \ jump} = - { { \alpha e^2 \lambda_c^2} \over
{2 \hbar} } n
\label{side1}
\end{equation}
where $\lambda_c= \hbar/mc=3.86
\times 10^{-11}$cm is the Compton wave length of free electron,
and $\alpha$ is the enhancement factor for the Bloch electrons
which is estimated to be around $10^4$. In terms of this
estimation one can estimate its ratio to the spin Hall conductance
$\sigma_s$ in Eq.~10 as
\begin{equation}
{ {\left| \sigma_{s }^{\rm side \ jump} \right|
} \over {\sigma_s}} \cong \alpha
(\lambda_c k_F)^2
\label{side2}
\end{equation}
which is around $10^{-3}$ even for the density as high as $n =
10^{20}$cm$^{-3}$. Therefore it is expected to be negligible. As
for the skew scattering contribution, it is at most of the order
of
\begin{equation}
{ {\left|\sigma_s^{\rm skew \ scattering }\right|} \over {\sigma_s} } \cong
\alpha (\lambda_c k_F)^2 \sqrt{ { { \varepsilon_F \tau_{p}} \over
{\hbar} } }
\label{skew}
\end{equation}
which is again much smaller than unity for the reasonable value of $\sqrt{{ {
\varepsilon_F \tau} \over {\hbar} }} \sim 10$. Therefore we have
the good reason to expect that our $\sigma_s$ in Eq.~10
gives the
dominant contribution, and hence the spin current estimated in
this paper can be observed in hole doped GaAs.

{}From Eqs.~\ref{side1}, \ref{side2} and \ref{skew}, we see
that the side jump and the skew scattering contributions scale
with the hole density like $n$ and $n^{4/3}$ respectively, while
our dissipationless spin conductivity $\sigma_s$ scales like
$n^{1/3}$. Therefore, one can plot the experimentally observed
total spin conductivity as $\sigma^T_s/n^{1/3}$ versus $n$, and
extrapolate to the limit of $n\rightarrow 0$. Only the
dissipationless spin conductivity contributes to the constant
intercept in this limit, and it can therefore be distinguished
from the impurity effects in an unique way.

\subsection*{Experimental detection of the spin current}

In detecting the spin current by attaching a ferromagnetic 
electrode as in Fig.~3{\bf A}, 
one should see a change of the electric current $I$
depending on the direction of
${\bf M}$.  For
$n=10^{19}{\rm cm}^{-3}$, the spin current density $j_{y}^{x}$ is
estimated as $ 2 e\hbar^{-1} j_{y}^{x} = (\sigma_s/\sigma) J_z \sim J_z$. 
The spin current 
$j_{y}^{x}$ induces the electric current $I$ 
when the magnetization of
the electrode is along $+x$-direction; no current is
observed for $-x$-direction, when the ferromagnetic electrode is 
a half-metal. In realistic situation,
the ratio of the electric currents $I(+x)/I(-x)$ is 
determined by
that of the tunneling probabilities for parallel and anti-parallel
spins at the interface with the electrode. This ratio is 
still
expected to be well larger than unity.

Another method of experimental detection of the spin current 
is to measure circular 
polarization of light emitted via recombination
with electrons as in Fig.~3{\bf B}. With
the steady electric current $J_z$ within the layer, the spin
current injected along the $y$-direction will be recombined with
the electrons supplied from the n-GaAs in the (In,Ga)As quantum
well. This can be regarded as the p-n junction with the opposite
voltage for opposite direction of the spin $s_x$. Therefore
the current along the circuit is different between up- and
down-spins with different direction, and hence the finite current
is observed for $I$. 

One can also detect the spin current by measuring the 
spins accumulated near the edges  of the
sample. 
The distribution of spin accumulation is determined by a balance between 
the spin current supply and the spin relaxation.
A diffusion equation for hole spins can be written as
\begin{equation}
\frac{\partial s_x(y,t)}{\partial t} - D\frac{\partial^2
s_x(y,t)}{\partial y^2} = -\frac{\partial j_{y}^{x}(y,t)}{\partial
y} - \frac{s_x(y,t)}{\tau_s}
\end{equation}
where $D=\mu k_{B}T/e$ is the diffusion constant. 
For simplicity, let us assume a semi-infinite sample extending for
$y<0$.
Considering a steady state and the spin current being
assumed to be $j_{y}^{x}(x) = j_{y}^{x} \theta (-y)$, we can solve the
above diffusion equation as
\begin{equation}
s_x(y) = j_{y}^{x} \sqrt{\frac{\tau_s}{D}} e^{\frac{y}{\sqrt{ D
\tau_s}}}.
\end{equation}
This solution means that the spins are distributed within the
length scale $L= \sqrt{ D \tau_s} $ with the total amount
$j_{y}^{x} \tau_s$ per unit area.
At room temperature, from $\tau_s= 100{\rm fsec}$ ({\it S8\/})
the length scale $L$ is estimated as $L\cong 4{\rm nm}$. We estimate
the upper bound for the electric current density as
$10^4$A/cm$^2$, and the corresponding
spin current $ 2 e j_{y}^{x} \sim 10^4$A/cm$^2$.
Therefore the spin density integrated over the depth is
given by $j_{y}^{x}\tau_s \sim 3\times 10^9 \mu_{{\rm B}}$cm$^{-2}$, 
too small to be observed optically.
However there are several ways to overcome this difficulty.
One is to lower the temperature. At $T=30$K,
$\tau_{s}$ reaches around $\tau_s \sim 30$psec ({\it S16\/}), nearly
300 times larger than that of room temperature.
and the total spin density  becomes
$\sim 10^{12}\mu_{{\rm B}}$cm$^{-2}$.
Another possibility is to inject the spin into
n-GaAs through the p-n junction. 
The spin lifetime of electrons in
GaAs is $100$psec, 
10$^3$ times longer than that of holes, 
and by suppressing the D'yakonov-Perel' mechanism
it could be even as long as $2{\rm nsec}$ ({\it S17\/}).
Therefore the accumulation depth $L$ becomes (0.1-1)$\mu$m,
comparable to the wavelength of the visible light,
with the area density $10^{12}$-$10^{13}\mu_{{\rm B}}$cm$^{-2}$
for the spins near the edges of the n-type GaAs.
Therefore it can be detected by the Kerr rotation
through surface reflection.

\subsection*{Supporting references and notes}

\begin{enumerate}
\item G. Dresselhaus, {\it Physical Review\/} {\bf 100}, 580 (1955).
\item 
E. I. Rashba,
{\it Soviet Physics, Solid State} {\bf 2}, 1109 (1960).
\item Y. A. Bychkov and E. I. Rashba.
{\it Journal of Physics C\/} {\bf 17}, 6039 (1984).
\item
J. M. Luttinger, {\it Phys. Rev.} {\bf 102}, 1030 (1956).
\item
A. Zee, {\it Phys. Rev. A} {\bf 38}, 1 (1988). 
Note that there is a misprint in this reference;
in the left column of page 2, the field strength
should be $F=imd\Omega$.
\item
M. V. Berry, {\it Proc. R. Soc. Lond. A} {\bf 392}, 45 (1984).
\item
   F. Meier, B.P. Zakharchenya
   {\it Optical Orientation}
   (North Holland, Amsterdam, 1984).
\item
   D. J. Hilton, C. L. Tang, {\it Phys. Rev. Lett.} {\bf 89}, 146601 (2002).
\item R. Karplus and J. M. Luttinger, {\it Physical Review\/} {\bf 95},
1154 (1954).
\item
J. Smit, {\it Physica\/} {\bf 21}, 877 (1955); {\it Physica\/} {\bf
24}, 39 (1958).
\item
W. Kohn and J. M. Luttinger, {\it Physical Review} {\bf 108}, 590
(1957).
\item
J. M. Luttinger, {\it Physical Review\/} {\bf 112}, 739 (1958).
\item
L. Berger, {\it Physical Review B\/}{\bf 2}, 4559 (1970).
\item
A. Cr{\'e}pieux and P. Bruno, {\it Physical Review B\/}{\bf 64}, 014416
(2001).
\item
   T. Jungwirth, Q. Niu, A. H. MacDonald,
   {\it Phys. Rev. Lett.}  {\bf 88}, 207208   (2002).
\item
S. Adachi, T. Miyashita, S. Takeyama, Y. Takagi, A. Tackeuchi,
{\it J. Lumin.} {\bf 72-74}, 307 (1997).
\item
Y. Ohno, R. Terauchi, T. Adachi, 
F. Matsukura, H. Ohno, {\it Phys. Rev. Lett.} {\bf 83}, 4196 (1999).
\end{enumerate}

\end{document}